\documentstyle[12pt,fleqn]{article}
\begin{document} 
\def\be{\begin{equation}} 
\def\ee{\end{equation}} 
\def\bea{\begin{eqnarray}} 
\def\eea{\end{eqnarray}} 
\def\beq{\begin{eqnarray*}} 
\def\eeq{\end{eqnarray*}} 
\def\ba{\begin{array}} 
\def\ea{\end{array}} 
\def\sss{\scriptscriptstyle} 
\def\ds{\displaystyle} 
\newcommand{\brho}{\mbox{\boldmath $\rho$}} 
\newcommand{\Bomega}{\mbox{\boldmath $\Omega$}} 
\def\e{{\bf e}} 
\def\v{{\bf v}} 
\def\r{{\bf r}}
\def\E{{\bf E}} 
\def\F{{\bf F}} 
\def\Fc{{\cal F}} 
\def\L{{\rm L}} 
\def\b{{\rm b}} 
\def\eff{{\rm eff}} 
\def\max{{\rm max}} 
\def\min{{\rm min}} 
\begin{center} {\large {\bf Stationary Velocity and Charge
      Distributions  of Grains in  Dusty Plasmas}\\ 
\vspace{0.2cm} A.G.~Zagorodny, P.P.J.M. Schram~$^*$,
S.A.~Trigger~$^{**}$, } 
\end{center} 
\begin{center} {\it Bogolyubov Institute for Theoretical Physics,
    National Academy  of Sciences of Ukraine\\
 14 B, Metrolohichna
    Str.,  Kiev 252143, Ukraine\\ 
$^*$ Eindhoven University of
    Technology\\ 
 P.O. Box 513, MB 5600 Eindhoven, The Netherlands\\
$^{**}$  Institute for High Temperatures, Russian Academy of
    Sciences\\  
13/19, Izhorskaya Str., Moscow 127412, Russia } 
\end{center} 
\vbox{\small Within the kinetic approach velocity and charge
  distributions of grains in  stationary dusty plasmas are calculated
  and the relations  between the effective temperatures of such
  distributions and plasma  parameters are established. It is found
  that the effective  temperature which determines the velocity grain
  distribution could  be anomalously large due to the action of
  accelerating ionic  bombarding force. The possibility to apply the
  results  obtained to the explanation of the increasing grain
  temperature in the course of the Coulomb-crystal melting by
  reduction of the gas pressure  is discussed. 
\vspace{0.2cm}
 
 {\em  This paper was received by Phys.Rev.Lett. on 11 August
  1999. As potential referees the authors offered to Editor the following
  persons: V.N.Tsytovich, Russia;  R.Bingham, UK; D.Resendes,
  Portugal; G.Morfill, P.Shukla, Y.M.Yu., Germany.}}

\vspace{0.4cm}
 Recently much attention has been payed to theoretical
  studies of various  problems of dusty plasma physics associated with
  grain dynamics  and grain charging (formation and melting of dusty
  crystals, influence  of charging on effective grain interaction,
  dust-acoustic wave  excitation, effect of grain charging on
  fluctuations and  electromagnetic wave scattering in dusty plasmas,
  etc.).  In such studies it is convenient to treat the grain charge
  as a new  variable (as was done for the first time in
  Ref. [1]). This makes  it possible to statistically describe the
  grain charge  distribution on equal footing with the spatial and
  velocity grain  distributions. Obviously, it is very important to
  know what are  the stationary (quasiequilibrium) grain distributions
  and what is  the relation of these distributions to plasma
  parameters. In spite  of the fact that statistical descriptions of
  dusty plasmas  have been already used in many papers, as far as the authors of this letter know neither grain
  charge, nor  velocity distributions for grains were studied within a
  consistent  kinetic approach. Usually, the problem is avoided by
  neglecting the  thermal dispersion of grain velocity and charge. In
  many cases  this is a rather reasonable approximation, but it could
  not be valid  when the properties of the grain subsystem and its
  dynamics are concerned.  

The purpose of the present paper is to
  describe stationary velocity  and charge distributions of grains in
  dusty plasmas in the case  of grain charging by plasma currents and
  to determine the  dependences of effective temperatures on plasma
  parameters. We study  dusty plasma consisting of electrons, ions,
  neutral molecules  and monodispersed dust particles (grains)
  assuming that every  grain absorbs all encountered electrons and
  ions. Such  collisions we define as \underline{charging collisions}.
  Collisions in which plasma particles do not touch the grain surface we call 
\underline{Coulomb elastic collisions}. Notice that the cross-sections of charging collisions are also determined by the Coulomb
  forces along with the  geometrical size of grains.

 Using the
  microscopic equations for  dusty plasmas and the relevant
  BBGKY-hierarchy [2] it is  possible to show that in the case of
  dominant influence of  charging collisions the kinetic equation for the
  grain distribution  function $f_g(X,t)\equiv f_g(\r,\v,q,t)$ ($q$ is
  the charge of  the grain) can be written as 
\bea
 &&\left\{\frac{\partial }{\partial t}+ \v\cdot
  \frac{\partial}{\partial \r} + {q\over m_g}\E\cdot \frac{\partial
  }{\partial  \v} \right\} f_g(X,t) = -\sum_{\sigma=e,i}\int d\v'
  \left[  \sigma_{g\sigma}(q,\v-\v')|\v-\v'| f_g(X,t) \right.\nonumber
  \\  &\!\!\!-\!\!\!& \left. \sigma_{g\sigma}(q-e_\sigma,
  \v-\v'-\delta\v_ \sigma)|\v-\v'-\delta\v_\sigma|
  f_g(\r,\v-\delta\v_\sigma,  q-e_\sigma,t)\right] f_\sigma(\r,\v',t),
  \eea 
 where $\sigma_{g\sigma}(q,\v)$ is the cross-section for charging: 

\be 
\sigma_{g,\sigma}(q,\v) =\pi a^2\left( 1-{2e_\sigma q\over m_\sigma
  v^2a}\right)\theta\left(  1-{2e_\sigma q\over m_\sigma v^2a}\right),
\ee
 $\theta(x)$ is the Heaviside step function, $a$ is the grain
  radius,  $f_\sigma(\r,\v,t)$ is the plasma particle distribution
  function  normalized by the particle density $n_\sigma$,
  $\delta\v_\sigma\equiv(m_\sigma/m_g)\v'$  is the grain velocity
  change due to the collision with  a plasma particle, subscript
  $\sigma$ labels plasma particle  species, the rest of the notations
  is traditional. Eq.~(1)  could be introduced also on the basis of
  physical arguments as was  done in Refs. [3,4]. In fact, the
  right-hand part of Eq.~(1)  describes the balance between the grains
  outcoming from the  phase volume element and those incoming to the
  same element due to  charging collisions. 

Taking into account the
  smallness of $e_\sigma$  and $\delta\v_\sigma$ it is possible to
  expand the right-hand  part of Eq.~(1) into a power series of these
  quantities. With  the accuracy up to the second order Eq.~(1) in the stationary isotropic and homogeneous case is reduced to 
\bea
 && \frac{\partial }{\partial \v}\left[ \frac{\partial }{\partial
     \v}\left(D_\parallel  f_g(\v,q)\right) +\beta\v f_g(\v,q)
   +\frac{\partial }{\partial  q}\left(q\gamma\v
     f_g(\v,q)\right)\right]\nonumber  \\ &\!\!\!+\!\!\!&
 \frac{\partial }{\partial q}  \left[\frac{\partial }{\partial
     q}\left( Qf(\v,q) \right) -I f_g(\v,q)\right]=0, 
\eea 
where $D_\parallel$, $\beta$, $Q$, $\gamma$ and $I$ are the
Fokker-Planck kinetic  coefficients generated by charging collisions
and given by 
\bea 
D_\parallel &\!\!\!\equiv\!\!\!& \sum_\sigma {1\over2} \left(
  {m_\sigma\over m_g}\right)^2 \int d\v' {(\v\cdot\v')^2\over v^2}
  |\v-\v'|\sigma_{g \sigma} (q,\v-\v') f_\sigma(\r,\v')\nonumber \\
  \beta &\!\!\! \equiv\!\!\!& \beta(q,v)=-\sum_\sigma {m_\sigma\over
    m_g} \int  d\v' {\v\cdot\v'\over v^2}
  |\v-\v'|\sigma_{g\sigma}(q,\v-\v')  f_\sigma(\r\v')\nonumber \\
  \gamma &\! \!\!\equiv\!\!\!& \gamma(q,v)=\sum_\sigma {m_\sigma\over
    m_g}  {e_\sigma \over q} \int d\v' {\v \cdot \v'\over v^2} |\v-\v'| \sigma_{g\sigma} (q,\v-\v') f_\sigma(\r,\v')\nonumber \\
Q &\!\!\!\equiv\!\! \!& Q(q,v)=\sum_\sigma {e_\sigma^2\over2} \int
d\v' |\v-\v'|  \sigma_{g\sigma}(q,\v-\v') f_\sigma(\r,\v')\nonumber \\
I &\!\!\!\equiv\!\!\!&  I(q,v)=\sum_\sigma e_\sigma\int d\v' |\v-\v'|
\sigma_{g\sigma}  (q,\v-\v') f_\sigma(\r,\v'). 
\eea 
The quantities $D_\parallel(q,v)$ and $Q(q,\v)$ characterize the grain
diffusion in the  velocity and charge space, respectively,
$\beta(q,\v)$ and $\gamma(q,\v)$  are the friction coefficients which
determine the  bombardment force $\F_\b(q,\v)=-m_g\beta(q,\v)\v$
associated with  charging collisions and the correction to this force
$\delta\F_\b(q,\v)=-m_g\gamma(q,\v)\v$ due  to the mutual influence of
the charge and velocity grain  distributions, $I$ is the grain
charging current.  Deriving the relation for $\beta(q,\v)$ we omit the
terms of  higher order in $(m_\sigma/m_g)$ associated with the tensor
nature of  the diffusion coefficient in velocity space (contribution of the transverse diffusion coefficient). With regard for the fact that $|I(q,v)/Q(q,\v)|\rightarrow \infty$ at
$e_\sigma\rightarrow 0$ and
$|\beta(q,\v)/D_\parallel(q,\v)|\rightarrow \infty$  at
$(m_\sigma/m_g)\rightarrow 0$, it is possible to  show that the
asymptotical solution of Eq.~(3) can  be written as 
\be 
f_g(\v,q)=n_{0g}Z^{-1} Q^{-1}(q,\v)e^{-W(q,v)+\lambda v^2}
D_\parallel^{-1}(q,\v)  e^{-V(q,v)+\varepsilon \delta q(v)}, 
\ee
 where
\bea 
W(q,v)  &\!\!\!=\!\!\!& -\int\limits_0^q dq/ {I(q',v)\over
  Q(q',v)}\nonumber  \\ V(q,v) &\!\!\!=\!\!\!& \int\limits_0^v dv'
{v'\over  D_\parallel(q,v')} \left\{
  \beta(q,v')+\frac{\partial}{\partial q}  (q\gamma(q,v'))
  -q\gamma(q,v')  \left[\frac{\partial W(q,v')}{\partial q}+
  \right. \right.  \nonumber \\ &\!\!\!+\!\!\!&
\left. \left. Q^{-1}(q,v')  \frac{\partial Q(q,v')}{\partial q}\right]
\right\}\nonumber \\  \delta q(v) &\!\!\!=\!\!\!& q-q(v), 
\eea 
$q(v)$ is the stationary charge of the grain moving with the velocity
$v$, given by the  equation 
\be 
I(q(v),v)=0, 
\ee 
$Z$ is a normalization constant, $\varepsilon$ and $\lambda$ are small functions. Substitution of
Eq.~(5)  into Eq.~(4) leads to 
\bea 
\varepsilon &\!\!\!=\!\!\!&
{1\over D_\parallel(q,v)}  \frac{\partial D_\parallel(q,v)}{\partial
  q} + \frac{\partial V(q,v)}{\partial q}\nonumber \\ \lambda
&\!\!\!=\!\!\!&  {1\over2v} \left\{ {1\over Q(q,v)}\frac{\partial
    Q(q,v)}{\partial v}  +\frac{\partial W(q,v)}{\partial v}
  +\varepsilon\frac{\partial  q(v)}{\partial v} \right\}. 
\eea
Eqs.~(5)--(8) give the  asymptotically exact solution of Eq.~(3) at
$(m_ge_\sigma/m_\sigma  q)\rightarrow \infty$. Further estimates
require the explicit form  of the kinetic coefficients. Assuming that
plasma particle  distributions are Maxwellian, one obtains the
following stationary  grain distribution with accuracy up to the
zeroth order in  $(qm_i/e_em_g)$: 
\be 
f_g(\v,q)=n_{0g}Z^{-1}
D_\parallel^{-1}(q,v)  e^{-{m_gv^2\over 2T_\eff (q)}} Q^{-1}(q,v)
e^{-{(q-q_0)^2 \over 2a\widetilde{T}_\eff}}, 
\ee 
where 
\bea 
T_\eff(q)
&\!\!\!=\!\!\!&  {2T_i(t+z)\over t-z+{(q-q_0)\over q_0} z[1+{t-z\over
    t+z}(1+{2Z_i\over 1+Z_i}  (1+t+z))]}\\[0.3cm] \widetilde{T}_\eff &\!\!\!=\!\!\!& {2\over
1+Z_i}\,  {1+t+z\over t+z} T_e, \eea and \bea D_\parallel (q,v)
&\!\!\!\simeq\!\!\!& D_0\left[ 1+{q-q_0\over q_0}\, {z\over
    t+z}\right]  \left( 1+{z\over t}\right)\nonumber \\ Q(q,v)
&\!\!\!=\!\!\!&  Q_0\left[ 1-{q-q_0\over q_0}\, {z(t+z-Z_i)\over
    (t+z)(1+Z_i)} \right] (t+z)(1+Z_i)\nonumber \\ D_0 &\!\!\!=\!\!\!&
{4\over3}  \sqrt{2\pi} \left({m_i\over m_g}\right) \left({T_i\over
    m_g}\right)  a^2n_iS_i \nonumber \\ Q_0 &\!\!\!=\!\!\!&
\sqrt{2\pi} \left(  {T_e\over T_i}\right) e_i^2 a^2n_iS_i, 
\eea
$n_{0g}$ is the  averaged number density of grains. Here, we use the
notation  
\beq 
z={e_e^2 Z_g\over aT_e}, \quad t={T_i\over Z_iT_e},
\quad S_i^2={T_i\over m_i},  \quad Z_g={q_0\over e_e}, \quad
Z_i=|{e_i\over e_e}|. 
\eeq 
The  quantity $q_0$ is the equilibrium
grain charge of  stationary particles satisfying the equation 
\be
I(q_0,0)=2\sqrt{2\pi}  a^2e_i^2n_iS_i \left[ 1+{z\over t}-
  \left({m_i\over m_e}\right)^{1/2}  \left( {T_e!
\over T_i}\right)^{1/2} {n_e\over Z_in_i}e^{-z}\right]=0. 
\ee 
For
typical values of  plasma parameters in dusty plasma experiments
$(t+z>1)$ and $Z_i=1$  we have: 
\beq 
\widetilde{T}_\eff\simeq
T_e. 
\eeq 
In such case  the thermal variation of the grain charge
$|q-q_0|^2$ is of  the order of $aT_e$ and $|q-q_0|z\sim
q_0\sqrt{e_e^2/aT_e}$. This  means that at weak plasma coupling
defined with the grain  size $(e_e^2/aT_e\ll 1)$ the effective
temperature of the grain  thermal motion $T_\eff(q)\equiv T_\eff$
reduces to 
\be  T_\eff\simeq 2T_i {t+z\over t-z} 
\ee 
and 
\beq
D(q,\v)\simeq D_0\left( 1+{z\over t}\right), \qquad Q(q,v)\simeq
Q_0(t+z)(1+Z_i). 
\eeq 
 Thus, in such case 
\be 
f_g(\v,q)={n_{0g}\over
  \sqrt{2\pi  a\widetilde{T}_\eff}} e^{-{(q-q_0)^2\over
    2a\widetilde{T}_\eff}}  \left( {m_g\over 2\pi T_\eff}\right)^{3/2}
e^{-{m_gv^2\over  2T_\eff}}. 
\ee 
This distribution describes the
equilibrium Maxwellian  velocity distribution and the Gibbs grain
charge distribution  with the temperatures $T_\eff$ and $\widetilde{T}_\eff$ respectively. In fact, the electric energy of charge variations of the electric capacity $a$ is equal
to $(q-q_0)^2/2a$  and thus, the charge distribution described by
Eq.~(15) can be interpreted  as an equilibrium distribution with
effective temperature  $\widetilde{T}_\eff$. At $t<1$, $z<1$ the
effective $\widetilde{T}_\eff$  exceeds the electron temperature. The
resulting velocity  distribution is described by the effective
temperature $T_\eff$. Even  in the case of neutral grains $(z=0)$ this
temperature is equal  to $2T_i$. The presence of the factor 2 is
associated with plasma particle  absorption by grains.

 Charging
collisions are inelastic and a  part of the kinetic energy of the ions
is transformed into  additional kinetic energy of the grains. This is
the difference between  the case under consideration and conventional
Brownian motion where  the velocity distribution is described by the
temperature of the  bombarding light particles. Eq.~(14) shows that
the effective temperature  of thermal grain motion could be anomalously high at $z\rightarrow t$. Physically it can be
explained by the decrease of the  friction coefficient with increase
of grain charge  
\beq
\beta(q,v)\simeq{2\over3}\sqrt{2\pi}\left({m_\sigma\over m_g}\right)
a^2n_iS_i \left(1-{z\over  t}\right)=\beta_0\left(1-{z\over t}\right)
\eeq 
The reason is that  the difference between the fluxes of ions
bombarding the grain surface  antiparallel to the grain motion and
parallel decreases with the  charge increase due to the specific
properties of the ionic  charging cross-section, which
charge-dependent part is larger  for ions moving with smaller relative
velocities (i.e.  in parallel direction). The condition $z=t$
corresponds to the zero  value of the friction force. 

Eq.~(3) and its
solutions (5),  (9), (15) were obtained under the assumption that the
Coulomb elastic  collisions could be neglected. In order to take
elastic collisions  into consideration Eqs.~(1), (3) should be
supplemented by the  appropriate collision terms, for example, by the Landau, or Balescu-Lenard collision integrals. 
We use the Balescu-Lenard collision integral in the Fokker-Planck form
which in  the case under consideration (isotropic spatially
homogeneous stationary  distribution) can be written as 
\be 
\left(
  \frac{\partial  f_g}{\partial t} \right)^C= \frac{\partial
  }{\partial \v}\cdot  \left[\frac{\partial }{\partial
    \v}(D_{\parallel C}(q,\v)  f_g(\v,q)) +\v \beta_C(q,\v)
  f_g(q,v)\right],  
\ee 
where $D_{\parallel C}(q,\v)$ and
$\beta_C(q,\v)$ are  the Fokker-Planck coefficients related to Coulomb
elastic  collisions (see, for example, [5], Chapter 8). With the
accuracy up to  the dominant logarithmic terms (in this approximation
Eq.~(16)  is reduced to the Landau collision term) such coefficients
can be  reduced to 
\bea 
D_{\parallel C}(q,\v) &\!\!\!\simeq\!\!\!&
{4\over3}  {\sqrt{2\pi}q^2\over m_g^2} \sum_{\sigma=e,i} {n_\sigma
  e_\sigma^2\over  S_\sigma} \ln \Lambda_\sigma \left(1-{v^2\over
    5S_\sigma^2} \right)\nonumber \\ \beta_C(q,\v)
&\!\!\!\simeq\!\!\!& {4\over3}  {\sqrt{2\pi}q^2\over m_g} \sum_{!
\sigma=e,i}\, {n_\sigma e_\sigma^2\over S_\sigma^3 m_\sigma} \ln
\Lambda_\sigma  \left(1-{v^2\over 5S_\sigma^2} \right), \quad
S_\sigma=\left(  {T_\sigma\over m_\sigma}\right)^{1/2}. 
\eea 
In
Eqs.~(16), (17)  we again neglect the contribution of the transverse
part of the  diffusion coefficient which gives a correction to
$\beta_C(q,v)$ of  higher order in $(m_\sigma/m_g)$ and we disregard
the grain-grain  Coulomb collisions, assuming the grain density to be
small  $(n_g<n_i(Z_i/Z_g)^2(S_g/S_i)^{1/2}(T_g/T_i))$. We introduced
also the Coulomb logarithms $\ln \Lambda_\sigma$ for each particle
species. Usually these quantities are estimated as $\ln \Lambda_\sigma
= \ln (k_{\max} / k_D)$, where $k_D=r_D^{-1}=(\sum(4\pi e_\sigma^2
n_\sigma /T_\sigma)^{1/2}$ and $k_{\max}$ is the inverse distance of
closest approach between colliding particles,
\be
k_{\max \sigma} \sim\frac{m_\sigma v^2}{\vert \epsilon_\sigma q \vert}
\sim \frac{3T_\sigma}{\vert \epsilon_\sigma q \vert} = r_{\L \sigma}^{-1}
\ee
($r_{\L \sigma}$ is Landau length). However, in the case of plasma
particle collisions with finite-size grains this estimate could be
invalid, since at $r_{\L \sigma} <a$ the Coulomb logarithm will include
the contribution of collisions with particles reaching the grain
surface, i.e. charging collisions.

An approximate modification of $\Lambda_\sigma$ is achieved by
treating $\ln \Lambda_\sigma$ as a logarithmic factor appearing in the
momentum transfer cross-section for Coulomb collisions. In the case of
finite size grains one obtains the following logarithmic factor
\bea
\ln \Lambda_\sigma = \ln \left( \sin \frac{\chi_{\max \sigma}}{2} / \sin
    \frac{\chi_{\min \sigma}}{2} \right), 
\nonumber
\eea
where $\chi_{\max \sigma}$ and $\chi_{\min \sigma}$
    are the scattering angles related to the minimal and maximal
    impact parameters  $b_{\min \sigma}$ and $b_{\max
          \sigma}$ by the Rutherford formula. Obviously,
        $b_{\min \sigma}$ should be determined from the
          condition that the distance of closest approach is equal to
          $a$ implying 
\be
b_{\min \sigma}=a \sqrt{1-\frac{2e_\sigma q}{m_\sigma v^2 a}}
  \theta (1-\frac{2e_\sigma q}{m_\sigma v^2 a}).
\ee

 Concerning the quantity $b_{\max \sigma}$, it is reasonable to put
   $b_{\max \sigma}=r_D +a$ instead of $b_{\max \sigma}=r_D$, since in
       the case of a finite size grain its screened potential is given
       by the DLVO-potential
\bea
\Phi(r)=\frac{q}{r} (1+\frac{a}{r_D})^{-1} e^{-(r-a)/r_D} ,
\nonumber
\eea  
rather than the Debye potential.\\
As a result we have
\bea
\ln \Lambda_i = \frac{1}{2} \ln
\frac {(r_D+a)^2+r_{\L i}^2}{(r_{\L i}+a)^2} 
\nonumber 
\eea

\bea
 \ln \Lambda_e = \frac{1}{2} \cases { {\ln
\frac {(r_D+a)^2+r_{\L e}^2} {(a-r_{\L e})^2}} & $a > 2r_{\L e}$ \cr
{\ln \frac {(r_D+a)^2+r_{\L e}^2} {r_{\L e}^2}} & $a < 2r_{\L e}$ \cr}
\eea
As is seen, at $r_{\L i}\gg r_D$  the ionic Coulomb logarithm can be a small quantity in
contrast to the case of  ideal plasmas. 

Comparing Eqs.~(3) and (16) it
is easy to see that  in order to take elastic Coulomb collisions into
account it is sufficient  to make the following replacements in the
obtained solutions  
\bea 
D_\parallel(q,v) &\!\!\!\rightarrow\!\!\!&
\widetilde{D}_\parallel(q,v)= D_\parallel(q,v) +D_{\parallel
  C}(q,v)\nonumber \\  \beta(q,v) &\!\!\!\rightarrow\!\!\!& \widetilde{\beta}(q,v)=\beta(q,v) +\beta_C(q,v). 
\eea 
In the case of
weak plasma  coupling $(e_e^2/aT_e\ll1)$ 
\bea
\widetilde{D}_\parallel(q,v)  &\!\!\!\simeq\!\!\!& D_0\left( 1+{z\over
    t} +{z^2\over t^2}  \ln\Lambda_i\right)\nonumber \\
\widetilde{\beta}(q,v)  &\!\!\!\simeq\!\!\!& \beta_0\left( 1-{z\over
    t}+2{z^2\over t^2}\ln\Lambda_i\right),  
\eea 
Thus, the correction
produced by the elastic collisions  could be of the same order as that
due to charging collisions.  The condition for dominant influence of
charging collisions  is 
\beq 
\left|1-{z\over t}\right| >2{z^2\over
  t^2} \ln\Lambda_i,  
\eeq 
which can be realized at small values of
$z/t$, or at $z|t\gg r_D^2/a^2(r_{\L i}\gg r_D^2/a)$.

Rigorously
speaking Eq.~(16) and thus Eq.~(22) are  definitely valid in the case
of weak coupling plasmas  $(r_{\L i}\ll r_D)$ since this is the
condition of the derivation  of the Balescu-Lenard (or, Landau)
collision term. However, it  is possible to expect that actually the
domain  of validity of Eqs.~(16), (22) is not too strongly restricted by such condition. This
assumption is in  agreement with the direct calculations of the
friction coefficient  (Coulomb collision frequency) in terms of the
binary collision  cross-sections. Beside that, as it was shown in
Ref. [6],  in the case of strong grain-plasma coupling the influence
of the  Coulomb collision is also small and the kinetic equation is
reduced  again to Vlasov equation. This means that fluctuation
evolution equations,  which solutions determine the explicit form of
the Balescu-Lenard  collision term, are the same as in the case of
weakly coupled plasmas  and thus Eq.~(16) continues to be valid.

 The
new kinetic  coefficients give the following effective temperature for
thermal  grain motion 
\be 
T_\eff =T_i {2\left(1+{z\over t}+{z^2\over
      t^2}\ln  \Lambda_i\right) \over 1-{z\over t} +2{z^2\over t^2}\ln
  \Lambda_i},  
\ee 
i.e. elastic collisions can produce a saturation of
the grain  temperature. However, in the case of dominant influence of charging collision $T_\eff$ can be still anomalously large. This fact can be used for a qualitative explanation of the
experimentally  observed grain temperatures which are usually much
higher than the  ion temperature, $T_g\gg T_i$ (see, for example
[7,8], $T_i\sim0.1$~eV,  $T_g\sim4\div40$~eV). Finally, we point out
that the obtained  results can be modified also for the case of a
plasma with a neutral  component. It is possible to introduce an
additional collision  term along with the term (16). Since the
collision integral  describing elastic collisions of neutrals with
grains also can  be represented in the Fokker-Planck form (it follows
from the  Boltzmann collision integral) the presence of neutrals
results in new  additions to $\widetilde{D}_\parallel$ and
$\widetilde{\beta}$,  namely 
\bea 
\widetilde{D}_\parallel(q,v)
&\!\!\!=\!\!\!&  D_0\left(1+{z\over t}+{z^2\over
    t^2}\ln\Lambda_i+{n_n\over n_i}  \left({m_n\over m_i}\right)^{1/2}
  \left({T_n\over T_i}\right)^{3/2} \right)\nonumber \\
\widetilde{\beta}(q,v)  &\!\!\!=\!\!\!& \beta_0 \left(1-{z\over t}+2{z^2\over t^2}\ln\Lambda_i +2{n_n\over n_i}
\left({m_n\over m_i} \right)^{1/2} \left({T_n\over T_i}\right)^{1/2}
\right).  
\eea 
As a result the effective temperature is modified into
\be 
T_\eff=2T_i  {\left(1+{z\over t}+{z^2\over
      t^2}\ln\Lambda_i+{n_n\over n_i}  \left({m_n\over
        m_i}\right)^{1/2}  \left({T_n\over T_i}\right)^{3/2}
  \right)\over  \left(1-{z\over t}+2{z^2\over t^2}\ln\Lambda_i
    +2{n_n\over n_i}  \left({m_n\over m_i}\right)^{1/2}
    \left({T_n\over T_i}\right)^{1/2} \right)}.  
\ee 
According to
Eq.~(25) the effective  temperature increases with decreasing neutral
density.  The influence of neutral density changes on the effective
temperature  would be especially important at $1-{z\over t}+2{z^2\over
  t^2}\ln\Lambda_i\stackrel{\sss<}{\sss\sim}0$.  In such a case a
decrease of the neutral gas pressure can produce  an anomalous growth
of $T_\eff$. That is in qualitative agreement  with the experimental
observation of melting of dusty crystals by  reduction of the gas
pressure [7,8]. 

The obtained results show that stationary  velocity
and charge grain distributions are  described by effective
temperatures different from those of the  plasma subsystem. These
effective temperatures are determined  by the competitive mechanics of
collisions: grain-neutral  collisions and elastic Coulomb collisions
result in the equalization  of the effective temperature to the
temperature of neutrals, or ions,  respectively, while charging
collisions can produce anomalous  temperature growth. That could be
one of the main mechanisms  of grain heating. 
\vspace{0.1cm} 
\noindent

This work was partially  supported by the Netherlands Organization of
Scientific Research (NWO)  and by the INTAS (grant 9600617). One of
the authors (A. Z.)  acknowledges support by NWO for his visit to
Eindhoven University of  Technology. 
\begin{description} 
\item{[1]} V.N.~Tsytovich, O.~Havnes, Comments Plasma
  Phys. Control. Fusion {\bf 15}, 267 (1995). 
\item{[2]} A.G.~Zagorodny, P.P.J.M.~Schram. S.A.~Trigger, to be published. 
\item{[3]} A.M.~Ignatov, J. Physique IV, {\bf C4}, 215 (1997). 
\item{[4]} S.A.~Trigger, P.P.J.M.~Schram, J. Phys. D.:
  Appl. Phys. {\bf 32}, 234 (1999). 
\item{[5]} S.~Ichimaru, Statistical Plasma Physics, Addison-Wesley,
  (1992). 
\item{[6]} X.~Wang, A.~Bhattacharjee, Phys. Plasmas {\bf 3}, 1189
  (1996). 
\item{[7]} A.~Melzer, A.~Homan, A.~Piel, Phys. Rev. E {\bf 53}, 3137
  (1996). 
\item{[8]} G.E.~Morfil, H.M.~Thomas, U.~Konopka, M.~Zuzic,
  Phys. Plasmas {\bf 5}, 1 (1999). 
\end{description} 
\end{document}